\documentclass{webofc}
\usepackage{hhline}
\usepackage{amsmath,amsfonts,amssymb}
\usepackage{bm}
\usepackage{aas_macros}
\usepackage{subcaption}
\usepackage{color}
\usepackage{multirow}
\usepackage{hhline}
\usepackage[varg]{txfonts}   
\usepackage{euclid}
\newcommand{\Om}{\Omega_\textrm{m}}

\begin{document}
\title{Galaxy cluster number counts with individual lensing mass estimates:  forecasts for \textit{Euclid}}
%
%

\author{E. Artis\inst{1}\fnsep\thanks{\email{emmanuel.artis@lpsc.in2p3.fr}} 
\and J-B Melin\inst{2}
  \and J. G. Bartlett\inst{2,3}
   \and Calum Murray\inst{1}, on behalf of the \textit{Euclid} Consortium} 

\offprints{----------, \email{----------}}

\institute{Univ. Grenoble Alpes, CNRS, Grenoble INP, LPSC-IN2P3, 53, avenue des Martyrs, 38000 Grenoble, France
 \and IRFU, CEA, Universit\'e Paris-Saclay, F-91191 Gif-sur-Yvette, France
  \and Universit\'e de Paris, CNRS, Astroparticule et Cosmologie,  F-75006 Paris, France
  \and Jet Propulsion Laboratory, California Institute of Technology, Pasadena, California, USA}

\abstract{%
  The $\Lambda$CDM model is gradually becoming challenged by observational data. Upcoming cosmological surveys will increase the number of detected galaxy clusters by several orders of magnitude. Therefore, clusters will shortly provide precise cosmological constraints and improve our understanding of structure formation in the Universe. 
  In the following, we present a cluster likelihood based on individual weak lensing mass estimates and forecast \Euclid 's performances within this framework.
   We use a matched filter for weak lensing mass estimation and model its characteristics with a set of simulations. We use the \textit{Flagship} N-body simulation to emulate the expected cluster mass distribution of a \Euclid-like sample and test our statistical framework against it. Finally, we simultaneously constrain the observable-mass relation and the cosmological parameters.}
\maketitle
\section{Introduction}
\label{intro}
Galaxy clusters are the most massive virialized structures in the universe.  Among the various ways of testing cosmological models with clusters \citep[e.g.,][]{2011ARA&A..49..409A}, their abundance as a function of mass and redshift is one of the primary probes.
The precision of current measurements is limited by our understanding of the observable-mass relation \citep[see][for a review]{2019SSRv..215...25P}. The upcoming years will transform our understanding of cluster cosmology, as the number of detected structures will increase by several orders of magnitude. Next-generation large surveys like \Euclid \citep{2016MNRAS.459.1764S} will provide an unprecedented quantity of data. 
In this proceeding, we propose an approach where a mass proxy is estimated for each cluster in the catalog to calibrate the observable mass relation. Individual masses have been introduced in cluster surveys \cite{2019ApJ...878...55B}, and  has been applied in the context of SZ observations by \cite{2019MNRAS.489..401Z}. 
The match filtering method extracting the lensing masses has been developed in Murray et al., in prep. 
In this work, we analyze the output of the matched filter and parameterize it to inject the results in a Poisson unbinned likelihood. 
We show that, in the context of a \Euclid -like survey, one can use this match filtering approach to constrain the observable-mass relation and the cosmological parameters jointly.

\section{Mass estimation}
\label{sec:mass_est}
Mass estimation is the current bottleneck of cluster abundance cosmology. For future surveys, we need to construct robust mass proxies with well understood properties.
In the case of \Euclid , the weak lensing measurements will provide the data required to estimate the dark matter halo mass. Clusters are primarily detected (in the optical) as overdensities of galaxies. One of the algorithms selected by the \Euclid consortium \citep{2019A&A...627A..23E} is \textit{AMICO} \citep{2018MNRAS.473.5221B}. It provides a natural mass proxy as an output, which is the  quantity related to the selection function. The weak lensing masses serve to calibrate its relation with the true halo mass.\\
In this section we briefly describe the two mass proxies that we use to forecast the performance of our framework.

\subsection{Lensing mass estimates}
\label{subsec:len_mass}
From Murray et al., in prep., the approach of the matched filter is the following: assuming a Navarro-Frenk-White (NFW) profile \cite{1996ApJ...462..563N}, we want to build an estimator of the scale radius $r_\mathrm{s}$ of a given cluster in the form
\begin{equation}
\label{eq:est_rs}
\tilde{r_\mathrm{s}}=\sum \Delta\Sigma_i\, h_i\; ,
\end{equation}
where $\Delta\Sigma_i$ is the measured excess surface density at a distance $r_i$ from the cluster center and $h_i$ are weights to be determined. From \cite{Wright_2000}, the theoretical excess surface density is proportional to the scale radius of the cluster  $\Delta\Sigma_{i,\mathrm{th}}=r_\mathrm{s}\tau_\mathrm{i}$. We thus have that $\Delta \Sigma_i = \Delta\Sigma_{i,\mathrm{th}} + n_i = r_\mathrm{s}\tau_i + n_i$, where $n_i$
is the noise, with the covariance matrix $\mathbf{S}=\mathbf{n}\mathbf{n}^\intercal$. If we want the filter to be unbiased and the noise to be minimal, a good choice of coefficients is
\begin{equation}
\label{eq:om_coef}
\mathbf{h}=\frac{\bm{\tau}^\intercal\mathbf{S}^{-1}}{\bm{\tau}^\intercal\mathbf{S}^{-1}\bm{\tau}}\; .
\end{equation}
We can then recover the mass $M_\mathrm{L}=4/3\pi\, \Delta\,(c_\Delta \tilde{r_\mathrm{s}})^3\,\rho_\mathrm{c}(z)$, where $c_\Delta$ is the concentration and $\rho_\mathrm{c}(z)$ is the critical overdensity of the universe at redshift $z$.
This estimator provides a robust summary statistic. Note that the lensing masses explicitly depend on the cosmological model. From the calculation of \cite{Wright_2000}, it can be shown that, within the framework of the matched filter and assuming that the clusters are at low to moderate redshifts ($z<1$ on average)
\begin{equation}
\label{eq:len_cosm}
M_\mathrm{L}(\Theta)\propto D_\mathrm{A}^{-3}(z|\Theta)\;\rho_\mathrm{c}^{-2}(z|\Theta)\; ,
\end{equation}
where $D_\mathrm{A}$ is the angular diameter distance and $\Theta$ is the set of parameters of the background cosmological model. For readability, we define the function $c(\Theta)\equiv D_\mathrm{A}^{-3}(z|\Theta)\; \rho_\mathrm{c}^{-2}(z|\Theta)$. Assuming that the masses where computed at a fiducial cosmology $\Theta_\mathrm{fid}$, we therefore have that the masses at a given cosmology $\Theta$ follow $M_\mathrm{L}(\Theta) = (c(\Theta)/c(\Theta_\mathrm{fid}))\,M_{\mathrm{L},\mathrm{fid}}$.
\subsection{Amplitude}
\label{subsec:amp}
Clusters are selected through their signature on the distribution of galaxies in the sky. In this work we focus on results from \textit{AMICO} \citep{2018MNRAS.473.5221B, 2019MNRAS.485..498M}.\\
%
%
%
This algorithm detects clusters as overdensities of galaxies in the sky, with no colour assumptions. The proxy of the mass is the normalization  of the cluster galaxy distribution, called the amplitude $A$. We assume a power law relation between the amplitude and the mass, $A(M)=A_0 \left(M/M_0\right)^\alpha$.
\section{Likelihood}
\label{sec:lik}
In this section, we briefly describe our likelihood model. We use an unbinned Poissonian likelihood, where the abundance of observed clusters per unit of lensing mass estimates $M_\mathrm{L}$, amplitude $A$ and redshift $z$ follows:
\begin{equation}
\label{eq:dist_a_ml}
\frac{\mathrm{d}\tilde{N}}{\mathrm{d}M_\mathrm{L}\,\mathrm{d}A\,\mathrm{d}z}(\Theta) =\Omega \int_{-\infty}^{+\infty}\mathrm{d} \ln M \;\frac{\mathrm{d}N}{\mathrm{d}\ln M\,\mathrm{d}\Omega\,\mathrm{d}z}(\Theta)\; f(A,M_\mathrm{L}|M,z,\Theta)\;
\chi(A(M),z)\; ,
\end{equation}
where $\mathrm{d}N/\mathrm{d}\ln M\,\mathrm{d}\Omega\,\mathrm{d}z$ is the halo mass function described in \cite{2016MNRAS.456.2486D}, $\chi(A(M),z)$ is the fraction of clusters detected at a given mass and redshift, $f(A,M_\mathrm{L}|M,z,\Theta)$ is the probability distribution function of the two observables $A$ and $M_\mathrm{L}$, at a given mass and redshift and $\Omega$ is the sky coverage of the survey.
In our case, we assume that,
\begin{equation}
\label{eq:pdf_obs_mass}
f(A,M_\mathrm{L}|M,z,\Theta)=\frac{1}{2\pi\,|\mathbf{C}|^{1/2}}
\exp \left[-\frac{1}{2}\left(\,\begin{pmatrix}
A\\
M_\mathrm{L}
\end{pmatrix}-\bm{\nu}(\Theta)\,\right)^\intercal \,\mathbf{C}^{-1}\, \left(\,\begin{pmatrix}
A\\
M_\mathrm{L}
\end{pmatrix}-\bm{\nu}(\Theta)\,\right)\right]\; ,
\end{equation}
with the correlation matrix,
\begin{equation}
\label{eq:cov_mat}
\mathbf{C}(M,z)=\begin{pmatrix}
\sigma_\mathrm{A}^2(M,z) & \rho\; \sigma_\mathrm{A}(M,z)\; \sigma_{M_\mathrm{L}}(M,z) \\
\rho\; \sigma_\mathrm{A}(M,z)\; \sigma_{M_\mathrm{L}}(M,z) & \sigma_{M_\mathrm{L}}^2(M,z)
\end{pmatrix}\; ,
\end{equation}
where $\rho$ is our correlation coefficient assumed to be constant and
\begin{equation}
\label{eq:nu_obs_bias}
\bm{\nu}(\Theta)=\begin{pmatrix}
A_0\, \left(\frac{(1-b_\mathrm{A})\,M}{M_0}\right)^\alpha\\
\frac{c(\Theta_\mathrm{fid})}{c(\Theta)}\,(1-b_\mathrm{L})\,M
\end{pmatrix}\; ,
\end{equation}
where $b_\mathrm{A}$ and $b_\mathrm{L}$ are potential biases, respectively on the amplitude and the lensing proxies.  The noise related to the amplitude $\sigma_\mathrm{A}(M,z)$ is parameterised following \cite{2019A&A...627A..23E}, while we assume that,
\begin{equation}
\label{eq:ml_siz}
\sigma_{M_\mathrm{L}}^2(M,z)=A_\mathrm{L}\, \overline{\theta_\Delta}^{\hbox{\hglue 0.9pt}\beta}(M,z)\,/\,n_\mathrm{gal}(z)+\sigma_\mathrm{int}^2(M,z)\,/\,n_\mathrm{gal}(z) \; ,
\end{equation}
where $n_\mathrm{gal}(z)$ is the number of background galaxies at redshift $z$, $\sigma_\mathrm{int}$ is the intrinsic scatter, and $\overline{\theta_\Delta}$ is the mean angular size of a cluster.
The fraction of detected objects is assumed to follow,
\begin{equation}
\label{eq:sel_fid}
\chi(A,z)=\mathbb{P}(q>q_\mathrm{obs})=\frac{1}{\sqrt{2\pi}}\int_{q_\mathrm{obs}}^{+\infty} \mathrm{d}q\,\exp\left(-\frac{(q-A/\sigma^2_\mathrm{A}(A,z))}{2}\right)\; ,
\end{equation}
where $q$ is the signal to noise ratio (S/N), and $q_\mathrm{obs}$ is the detection treshold.

\section{Simulation and mock generation}
\label{sec:sim_moc}

\begin{figure}
   \centering
   \includegraphics[width=0.7\hsize]{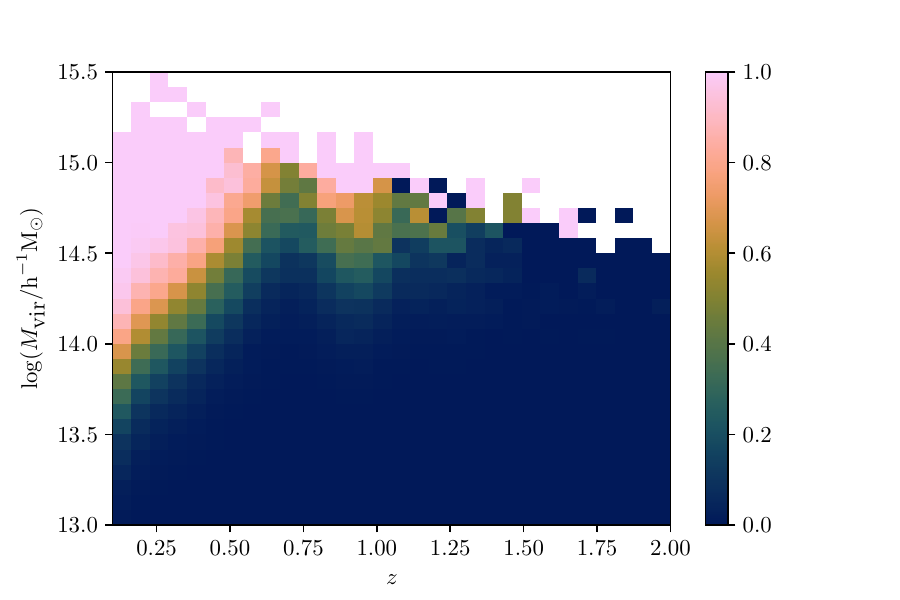}
      \caption{Distribution of the clusters selected in the Flagship dark matter halo catalog. We set the S/N ($q_\mathrm{obs}=6$, from equation \ref{eq:sel_fid}), so that we remove most of the low mass systems for which the uncertainties are high, especially regarding the selection function. The color scale represents the fraction of detected clusters.}
         \label{fig:cluster_sel}
   \end{figure}

The impact of the systematics that we will have to face must be measured at the precision of a \Euclid -like survey, which means $15,000 ~\mathrm{deg}^2$, $z\in [0,2]$, reliable lensing information and baryonic physics correctly incorporated.
This represents an enormous volume, and currently, no data set possesses these characteristics.
To tackle this issue, we have chosen to use three different simulations, each one having the required properties that we are looking for. The \textit{RayGalGroupSims} \citep{2019MNRAS.483.2671B}, are used to derive the lensing properties. For the selection function, and the performance of the detection algorithm, we refer to the the results of  \citep{2019A&A...627A..23E}. Finally, we used the \textit{Flagship} dark matter halo catalog, the official \Euclid simulation, as a baseline to reconstruct a mock survey that is as close as possible from a \Euclid-like data set.
The \textit{RayGalGroupSims} are a set of N-body simulations that are dedicated to ray-tracing. The main interest of these simulations is their spatial resolution ($5 \,h^{-1}\mathrm{kpc}$), and the excellence of the lensing calculations.
We use these simulations to test and constrain the relation on the scatter of the lensing mass estimates introduced with equation \ref{eq:ml_siz}. They extend up to $z=0.5$, so we assume that our results are valid for higher redshifts. Precisely, we fit the normalization and the power law of the relation. In this study, we fix the value of the intrinsic scatter $\sigma_\mathrm{int}$ to 0.3. We do verify that our fiducial value corresponds to a reasonable fit of the masses. The results are presented in table \ref{tab:all_params}.\\
%
  %
%
%
In order to create a catalog of objects similar to the one expected from \Euclid, we select a spherical region of 15,000~$\mathrm{deg}^2$, in which we select all the halos down to a virial mass of $M_\mathrm{vir, cut} = 10^{13} ~ h^{-1}\textrm{M}_\odot$.
Starting from these masses, we assign to each halo an amplitude, $A$, and a lensing mass $M_\mathrm{L}$ using the distribution presented in section \ref{sec:lik}.
From our dark matter halo catalog, we select the objects that would be detected by \textit{AMICO} according to equation \ref{eq:sel_fid}. We fix the S/N threshold in agreement with the expected number of objects to be detected with \Euclid. 
We chose an S/N threshold of $q_\textrm{obs}= 6$,
which gives provides us with a catalog of 56,549 clusters. 
The clusters' distribution is represented in figure \ref{fig:cluster_sel}.
 %
%
%

\section{Performance forecasts}
\label{sec:per_for}
\begin{table*}
\caption{True values adopted for the parameters, priors and results with $1\sigma$ uncertainties. Some parameters are left blank in the posterior section when they are not constrained. The dark energy equation of state parameters, are in \textcolor{blue}{blue}, with the following parameterization: $w(a)=w_0+w_\textrm{a}(1-a)$ \citep[]{CHEVALLIER_2001}. The true values, for the lensing mass estimates are the best fit values introduced in equation \ref{eq:ml_siz}. Note that although it is present in our table, the intrinsic scatter is not fitted, but fixed. As expected, the value of the power $\beta$ is close to the intuitive value $-2$, as the errors should be proportional to the inverse of the angular area covered by the cluster.}
\label{tab:all_params}
\centering
\scalebox{0.68}{
\begin{tabular}{ c | c | c | c | c}
  Parameter & Meaning & True value & Prior & Posteriors\\
  \hline
  \multicolumn{5}{c}{Cosmological parameters}  \\
  \hline
  $\Omega_\textrm{m}$ & Matter density & 0.319 & [0.1,0.5] & $0.322\pm 0.0019$ \\
  $\sigma_8$     &  r.m.s. matter fluctuation & 0.83 & [0.6,0.9] & $0.835\pm 0.0056$\\
  $H_0$       & Hubble constant ($\mathrm{km}\cdot\mathrm{s}^{-1}\cdot\mathrm{Mpc}^{-1}$) & 67 & $\mathcal{N}(67, 1000)$ & $65.28\pm 1.21$\\
  $\Omega_\textrm{b}$ & Baryonic density & 0.049 & $\mathcal{N}(0.049, 0.0026)$ and $\Omega_\textrm{b} >0$ & - \\  
  $n_\mathrm{s}$       &Spectral index & 0.96 & [0.87,1.07] & - \\
  \textcolor{blue}{$w_0$} & \textcolor{blue}{First parameter} &\textcolor{blue}{$-1$} & \textcolor{blue}{$[-2.5,0]$} &\textcolor{blue}{$-0.967\pm 0.034$}\\
  \textcolor{blue}{$w_\mathrm{a}$} & \textcolor{blue}{Second parameter} & \textcolor{blue}{0} & \textcolor{blue}{$[-2,2]$} & \textcolor{blue}{$0.19\pm 0.12$}\\
  \hline
  \multicolumn{5}{c}{Halo Mass Function parameters}  \\
  \hline
  $a_0$ & High mass cut-off & 0.938 & \multirow{4}{*}{Covariance matrix fitted on the \textit{Flagship}} & -\\
  $a_\textrm{z}$     & Redshift dependence $a_0$ & $-0.12$ & & -\\
  $p$       & Shape at low masses & $-0.2$ & & -\\
  $A_{0,\textrm{MF}}$  & Normalization & 0.31 & & -\\  

  \hline
  \multicolumn{5}{c}{Richness-mass parameters} \\
  \hline
  $A_0$       & Normalization & 0.5 & [0.05,5] & $0.495\pm 0.0045$\\
  $\alpha$      & Slope & 0.5 & [0.05,2] & $0.5\pm 0.0012$\\
  \hline
  \multicolumn{5}{c}{Lensing mass estimates uncertainties} \\
  \hline
  $A_\mathrm{L}$       & Normalization of the scatter & 0.836 & [0.001,4] & $0.758\pm 0.042$\\
  $\beta$     & Slope & $-2.09$ & $[-6,0]$ & $-2.2\pm 0.136$\\  
  $\sigma_\textrm{int}$     & Intrinsic scatter & 0.3& [0.001,2] & $0.35\pm0.025$\\  
  \hline
  \multicolumn{5}{c}{Correlation richness-lensing mass estimates}  \\
  \hline
  $\rho$       & - & 0.6 & [0,1] & $0.599\pm 0.0037$\\
  \hline
  \multicolumn{5}{c}{Bias on the richness} \\
  \hline
  $b_\textrm{A}$       & - & 0 & [0,0.99] & -\\
  \hline
  \multicolumn{5}{c}{Bias on the lensing masses} \\
  \hline
  $b_\textrm{L}$       & - & 0 & Fixed  & -\\
  \hline
  
 \end{tabular}
 }
  \end{table*}

\begin{figure}[h]
\centering 
\begin{minipage}{.49\textwidth}
  \centering
  \includegraphics[width=\linewidth]{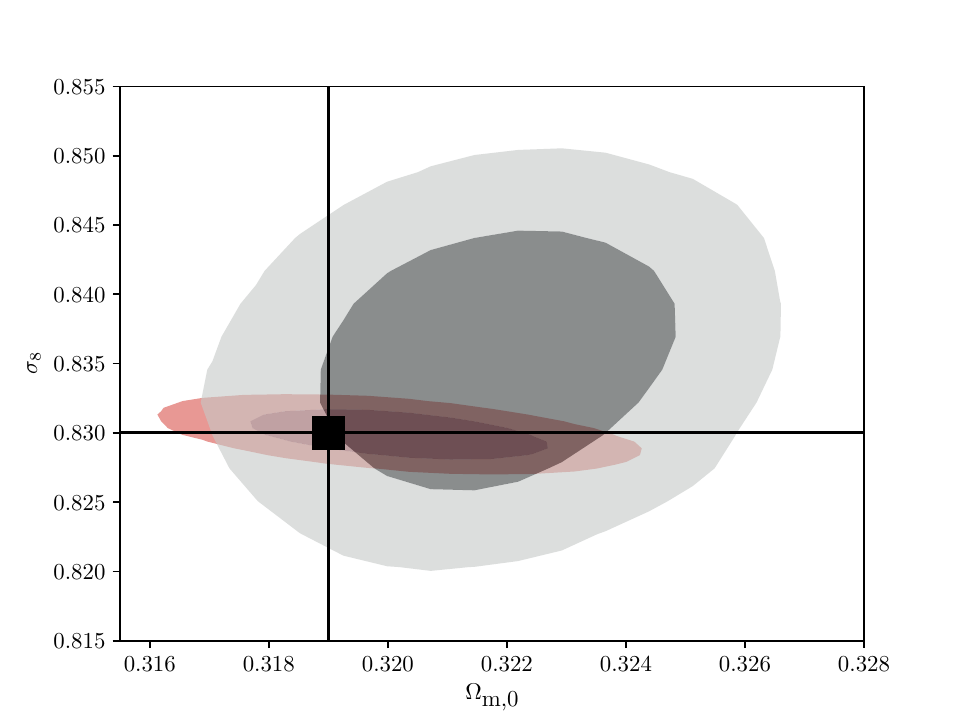}
\end{minipage}
\begin{minipage}{.49\textwidth}
  \centering
  \includegraphics[width=\linewidth]{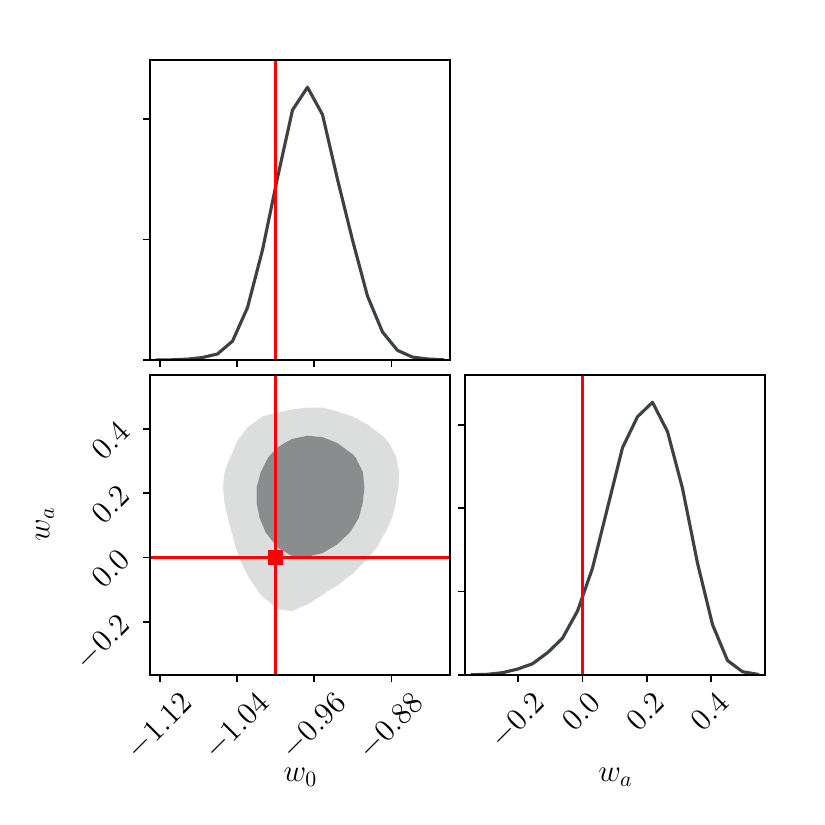}
\end{minipage}

\caption{\textbf{Left}: 68 and 95\% constraints obtained with our analysis, in case of a $\Lambda$CDM cosmology. The grey ellipse is the realistic case for which all the identified sources of systematics listed in table \ref{tab:all_params} are free. With the \textcolor{red}{red} ellipse , we also provide a reference case, for which all the systematics are fixed. \textbf{Right}: Constraints obtained on the dark energy equation of state parameters. The position of the best fit compared to the confidence ellipses in the grey case is an expected feature that results from the statistical production of the catalogs. }
\label{fig:cosmo_params}
\end{figure}

   We first consider a standard $\Lambda$CDM cosmology. The full set of fitted parameters is described in table~\ref{tab:all_params}.
   Figure \ref{fig:cosmo_params} shows the $\Om - \sigma_8$ plane for the cases when all the systematics are entirely known and when we freed all the nuisance parameters. As illustrated, it is clear that a joint fitting of the mass-observable relation and cosmological parameters is possible and provides competitive results, providing that one has statistically robust lensing mass estimates.
   %
   %
 Then we can compare these results to the one presented by \cite{2016MNRAS.459.1764S}. Our contours are comparable to what they are presenting for their number counts only analysis. Figure \ref{fig:cosmo_params} shows the results obtained for a $w_0w_a$CDM cosmology%

\section{Conclusion}
In this work, we presented a method designed for the analysis of \Euclid's galaxy cluster sample. This framework consists in the computation of individual lensing mass estimates, for each cluster in the catalog.
We use numerical simulations to reproduce the expected properties of \Euclid's cluster sample. Our final cluster catalog is based on the \textit{Flagship} DM halo catalog and comprises approximately $6 \times 10^4$  objects. We model the cosmological dependence of the lensing mass estimates, which is taken into account in the final estimation. We show that it is possible to calibrate the cosmological parameters and the parameters of the observable-mass relation jointly.
Our analysis demonstrates the potential of individual masses analyses in the context of large optical surveys. We assumed analytical expressions  for the selection function, observable-mass relation, and cosmological dependence of the lensing mass estimates, relying on a set of numerical simulations. Each one of these elements has to be refined for the final analysis and is under investigation. %
\section*{Acknowledgements}
\AckECon
%
%
%
\bibliography{references}
%
%

\end{document}